\documentclass[
 reprint,
 amsmath,amssymb,
 aps,
]{revtex4-2}
\usepackage{comment}
\usepackage[normalem]{ulem}
\usepackage{graphicx}
\usepackage{dcolumn}
\usepackage{float}
\usepackage{bm}
\usepackage{hyperref}
\usepackage{dirtytalk}
\usepackage{xcolor}
\usepackage{soul}
\usepackage{amssymb}
\usepackage{lipsum}
\usepackage{comment}
\usepackage{graphicx}
\usepackage{dcolumn}
\usepackage{bm}
\usepackage{hyperref}
\usepackage{siunitx}
\usepackage{dirtytalk}
\hypersetup{
    colorlinks=true,
    linkcolor=blue,
    filecolor=magenta,      
    urlcolor=cyan,
    pdftitle={Overleaf Example},
    pdfpagemode=FullScreen,
    }

\begin{document}
\preprint{APS/123-QED}
\title{Emergent Rashba spin-orbit coupling in bulk gold with buried network of nanoscale interfaces}
\author{Shreya Kumbhakar$^{1,\dagger, \ast}$}
\author{Banashree Debnath$^{1,\dagger, \ast}$}
\author{Tuhin Kumar Maji$^{1}$}%
  \author{Binita Tongbram$^1$}%
   \author{Shinjan Mandal$^1$}
  \author{T. Phanindra Sai$^1$}%
  \author{T.V. Ramakrishnan$^1$}
 \author{Manish Jain$^1$}
 \author{H. R. Krishnamurthy$^{1,2}$}
  \author{Anshu Pandey$^3$}%
 \author{Arindam Ghosh$^{1,\ast}$}%
\affiliation{%
 $^1$Department of Physics, Indian Institute of Science, Bangalore 560012, India
}%
\affiliation{%
 $^3$Solid State and Structural Chemistry Unit, Indian Institute of Science, Bangalore 560012, India
}
\affiliation{$^2$International Centre for Theoretical Sciences, Tata Institute of Fundamental Research, Bangalore 560089, India}
\begin{abstract}
The Rashba effect, which plays a crucial role in fundamental materials physics and potential spintronics applications, has been engineered in diverse systems, including semiconductor quantum wells, oxide heterostructures, metallic surfaces, topological insulators, ferroelectrics, etc. However, 
generating it in systems that preserve bulk inversion symmetry (BIS), for example, in bulk metals, has not been possible so far. We demonstrate a unique strategy to introduce and tune Rashba spin-orbit interaction (SOI) to unprecedented magnitudes in inversion-symmetric solids, by incorporating ultra-small silver nanoparticles in bulk gold. The near-identical lattice constants of Ag and Au allowed dense packing of the Ag/Au hetero-interfaces without compromising the global BIS. By varying the density of embedded nanoparticles, we generate Rashba SOI in a bulk metal with coupling strength $\sim15$ meV.\r{A}, higher than any known system preserving BIS globally and up to $\sim20$ times increase in the spin-orbit scattering rate. We argue that the combined effect of charge-transfer at the interfaces and polaronic localization enhances the SOI.
\end{abstract}
\maketitle
\noindent
\subsection*{Introduction}
Materials lacking inversion symmetry exhibit Dresselhaus or Rashba spin-orbit coupling (SOC), which lifts the spin degeneracy of the bands without an external magnetic field.
Bulk inversion asymmetry (BIA) in three-dimensional bulk periodic solids, like non-centrosymmetric zinc-blende structures, leads to the Dresselhaus SOC \cite{PhysRev.100.580}, whereas Rashba SOC emerges from structural inversion asymmetry (SIA), like in two-dimensional surfaces or interfaces \cite{1984JETPL..39...78B,manchon2015new}. 
Intensive efforts have been employed over several decades to engineer artificial heterostructures from metals, semimetals, semiconductors, and insulators that show Rashba coupling \cite{manchon2015new,bihlmayer2022rashba}, which is modelled by the Bychkov-Rashba Hamiltonian $H_R=\alpha_R\vec{\sigma}. (\vec{k}\times \hat{z})$. Here, $\vec{\sigma}$ and $\vec{k}$ are the electron's spin and wavevector, respectively; $\alpha_R$ is the Rashba parameter, and $\hat{z}$ is a unit vector along the direction of SIA \textit{i.e.} perpendicular to the surface/interface. 
The magnitude of $\alpha_R$ ranges from $\sim 4-6$ meV.\r{A} in III-V semiconductor quantum wells \cite{PhysRevB.74.033302,PhysRevB.89.205201} to $\sim10-50$ meV.\r{A} in interfaces of complex oxides heterostructures \cite{PhysRevLett.104.126803,lin2019interface} to $\sim30-3000$ meV.\r{A} on metallic surfaces or interfaces \cite{PhysRevLett.77.3419,PhysRevB.66.245419,PhysRevLett.98.186807,marchenko2012giant}. 
An ongoing quest over the years is the search for materials showing \textit{bulk} Rashba effect, which had been mostly associated with 2D surfaces or interfaces \cite{bihlmayer2022rashba,zhang2014hidden}. 
This was encouraged after the discovery of giant bulk Rashba splitting in polar semiconductors like BiTeX (X=Br, Cl, or I) \cite{ishizaka2011giant,bahramy2011origin,https://doi.org/10.1002/adma.201203199}, GeTe, SnTe, and organometal halide perovskites \cite{doi:10.1126/sciadv.1700704}.
However, bulk systems showing the Rashba effect discovered so far lack a centre of inversion globally possessing a non-centrosymmetric structure and hence are observed in a restricted material domain~\cite{zhang2014hidden}. Because of intrinsic structural symmetry, the Rashba interaction is absent in most metals, which are important for numerous spin-manipulation-related applications~\cite{wang2016giant,amin2018interface,miron2011perpendicular,lee2016emerging}. 
This poses a fundamental question of whether it is possible to induce Rashba physics in a bulk system with global inversion symmetry.
\begin{figure*}[ht]
    \centering
    \includegraphics[width=1\linewidth]{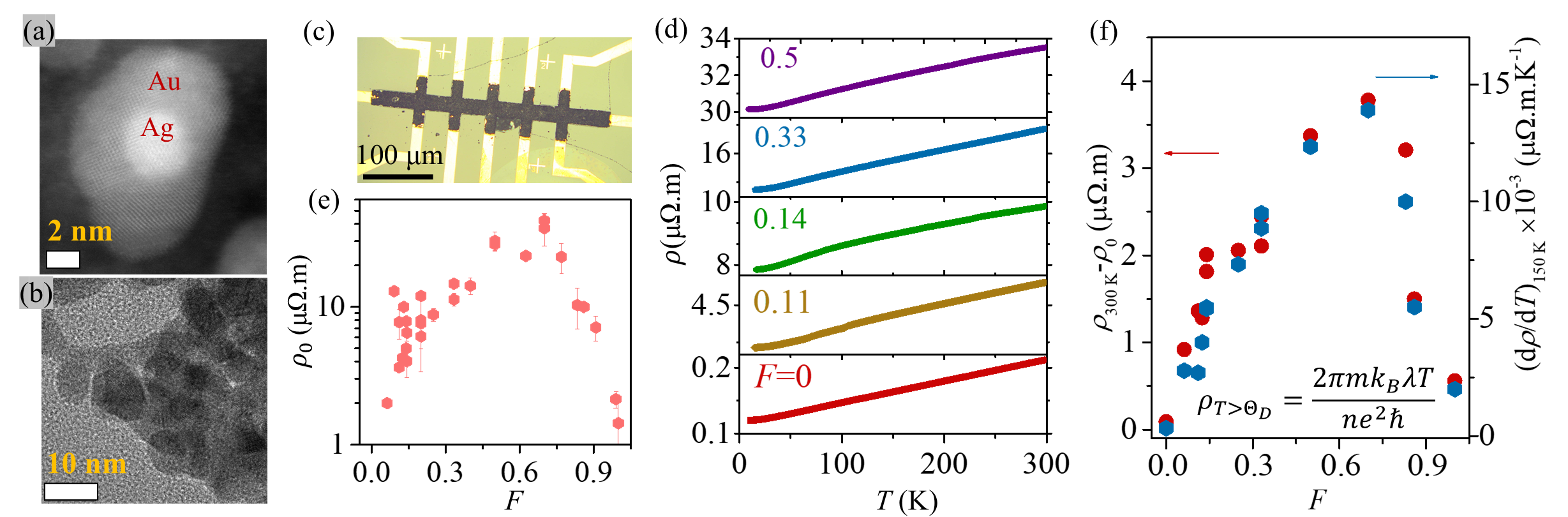}
    \caption{\textbf{Structural and electrical characterization of Ag@Au nanohybrid films}: (a) z-contrast variation in High-Angle Annular Dark-Field imaging (HAADF) of Ag@Au nanohybrid indicates a spherical Ag nanoparticle ($\sim2$~nm dia), represented by the brighter white region, is hosted inside the Au lattice, represented by the darker grey region. (b) A High-Resolution Transmission Electron Microscopy (HRTEM) image shows multiple Ag nanoparticles (AgNPs) of high crystallinity embedded inside a crystalline matrix of Au. (c) shows the typical image of a film, represented by the darker region deposited on pre-patterned Cr/Au electrodes on a glass substrate, indicated by gold-coloured regions. (d) $T$-dependence of resistivity, $\rho$ for films with varying Ag-Au volume fraction,  $F=V_{\mathrm{Ag}}/(V_{\mathrm{Ag}}+V_{\mathrm{Au}})$.
(e) Variation of the residual resistivity, $\rho_0$ (defined as the resistivity, $\rho$ at the base temperature $T\sim6$~K) with $F$. 
(f) The $T$-dependent component of resistivity, $\rho_{300~\mathrm{K}}-\rho_0$, and the slope of $\rho-T$ at $T\sim150$~K are plotted in the left and right axes, respectively, for different values of $F$. 
    }
    \label{fig1}
\end{figure*}

Fundamentally, the Rashba interaction derives from the electric field associated with an interfacial potential gradient that is transformed to a magnetic field, which relativistically couples to the electron's spin. The utilization of interfaces has hence been a potential strategy to induce the Rashba effect. This has been demonstrated even on a metallic platform, where a non-centrosymmetric artificial superlattice of [Pt/W/Co]$_N$ ($N=10$) enables Rashba physics in bulk metallic systems~\cite{ham2023bulk}, with $\alpha_R$ as large as $\sim12$~meV.\r{A}. However, not only is the $\alpha_R$ strictly limited by the thickness of the metallic layers, but also its scalability to the bulk limit (\textit{i.e.} large $N$) is experimentally challenging and yet to be demonstrated. Consequently, engineering the Rashba effect in a truly bulk metal that also preserves the bulk inversion symmetry remains unresolved.

In this work, we fabricate a crystalline matrix of Au that encloses multiple ultra-small Ag nanoparticles of diameter $\sim2$ nm~\cite{doi:10.1021/acsaelm.3c00379}.
By performing quantum transport measurements, we show that the hybrid system exhibits a Rashba SOC with $\alpha_R\sim 15$ meV.\r{A}. Our experiments suggest that this effect arises from the breaking of structural inversion symmetry and the strong dipole field generated locally at the Ag/Au interfaces because of the difference in the onsite electrochemical potentials of Au and Ag, which also leads to strong effective electron-phonon coupling~\cite{kumbhakar2025engineering}, as well as a soft energy gap in the electronic spectrum at low energies.  
The Rashba coupling could be tuned over a factor of $\sim 20$ by varying the density of nanoparticles, thereby demonstrating a novel strategy to achieve tunable Rashba interaction in a bulk noble metal by embedding a network of nanoscale interfaces. 
\subsection*{Results}
\subsubsection*{Structural characterization}
The Ag@Au nanohybrids are synthesized with a colloidal co-precipitation process \cite{doi:10.1021/acsaelm.3c00379,kumbhakar2025engineering} that allows exceptional tunability in the density of the Ag/Au interfaces while retaining a global inversion symmetry owing to the near-perfect lattice matching of Ag and Au. \textcolor{black}{(See Supplementary Information (\textbf{SI}) Section I for the details on synthesis, which are reproduced following references~\cite{doi:10.1021/acsaelm.3c00379,kumbhakar2025engineering})}.
Fig.\hspace{0.1cm}\ref{fig1}\textcolor{blue}{(a)} shows a typical z-contrast of a High-angle annular dark-field (HAADF) image of the Ag@Au nanohybrid, where the darker crystalline region of Au hosts a dispersion of the brighter region of Ag of diameter $\sim2$~nm. 
Fig.~\ref{fig1}\textcolor{blue}{(b)} shows the High-Resolution Transmission Electron Microscope (HRTEM) image of a section of the crystalline Au matrix embedding multiple Ag nanoparticles.\cite{doi:10.1021/acsaelm.3c00379,kumbhakar2025engineering}.It is also important to note from the HRTEM image that the spherical nature of the AgNPs and the Ag/Au interface is retained in the embedded structure. (See \textbf{SI} Section II for HRTEM images at different values of $F$). Our synthesis protocol offers precise tunability over the average radii of the embedded Ag nanoparticles (AgNPs), $r_{\mathrm{Ag}}$, and their volume fraction, $F=V_{\mathrm{Ag}}/(V_{\mathrm{Ag}}+V_{\mathrm{Au}})$. For all measurements reported in this manuscript, $r_{\mathrm{Ag}}$ is fixed at $\sim1$~nm.  
Hence, the Ag-filling fraction, $F$, is also proportional to the volume density, $F/r_{\mathrm{Ag}}$, of the interfaces. The as-synthesized nanohybrids are then assembled on pre-patterned Cr/Au electrodes on a glass substrate with chemical `cross-linking' protocols to form a film with average thickness $\sim3$~$\mu$m for electrical transport measurements \cite{doi:10.1021/acsaelm.3c00379} (See details on film preparation in \textbf{SI} Section III). A typical optical image of a patterned film is shown in Fig.~\ref{fig1}\textcolor{blue}{(c)}, where the dark region corresponds to the Ag@Au nanohybrids (NHs). 
\subsubsection*{ Temperature-dependent Resistivity measurements}
We have performed four-probe resistance measurements to estimate the resistivity, $\rho$, of the films. 
Fig.~\ref{fig1}\textcolor{blue}{(d)} shows the temperature ($T$)-dependence of $\rho$ for films with varying $F$. All the films are metallic down to $T\sim6$~K, indicating the formation of a true metallic composite from the Ag@Au NH that is devoid of tunnel barrier or insulating chemical residues across the Ag/Au interfaces. Fig.~\ref{fig1}\textcolor{blue}{(e)} shows the residual resistivity, $\rho_0$ ($\rho$ at temperature $\sim6$~K) for different Ag@Au NH films. We observe $\rho_0$ to increase with increasing $F$, reaching $\sim30~\mu\Omega$.m at $F\sim0.7$, which is two orders of magnitude higher compared to that of pristine Au ($\sim0.1~\mu\Omega$.m). Such a significant enhancement arises from the buried Ag/Au interfaces that form the dominant source of scatterers and increase the resistivity almost linearly with increasing interface density, $F/r_{\mathrm{Ag}}$~\cite{doi:10.1021/acsaelm.3c00379}. By increasing $F$ further, $\rho_0$ decreases, approaching $\sim1~\mu\Omega$.m for a pure Ag nanoparticle (AgNP) film ($F=1$).
 \begin{figure*}[ht!]
    \centering
    \includegraphics[width=1\linewidth]{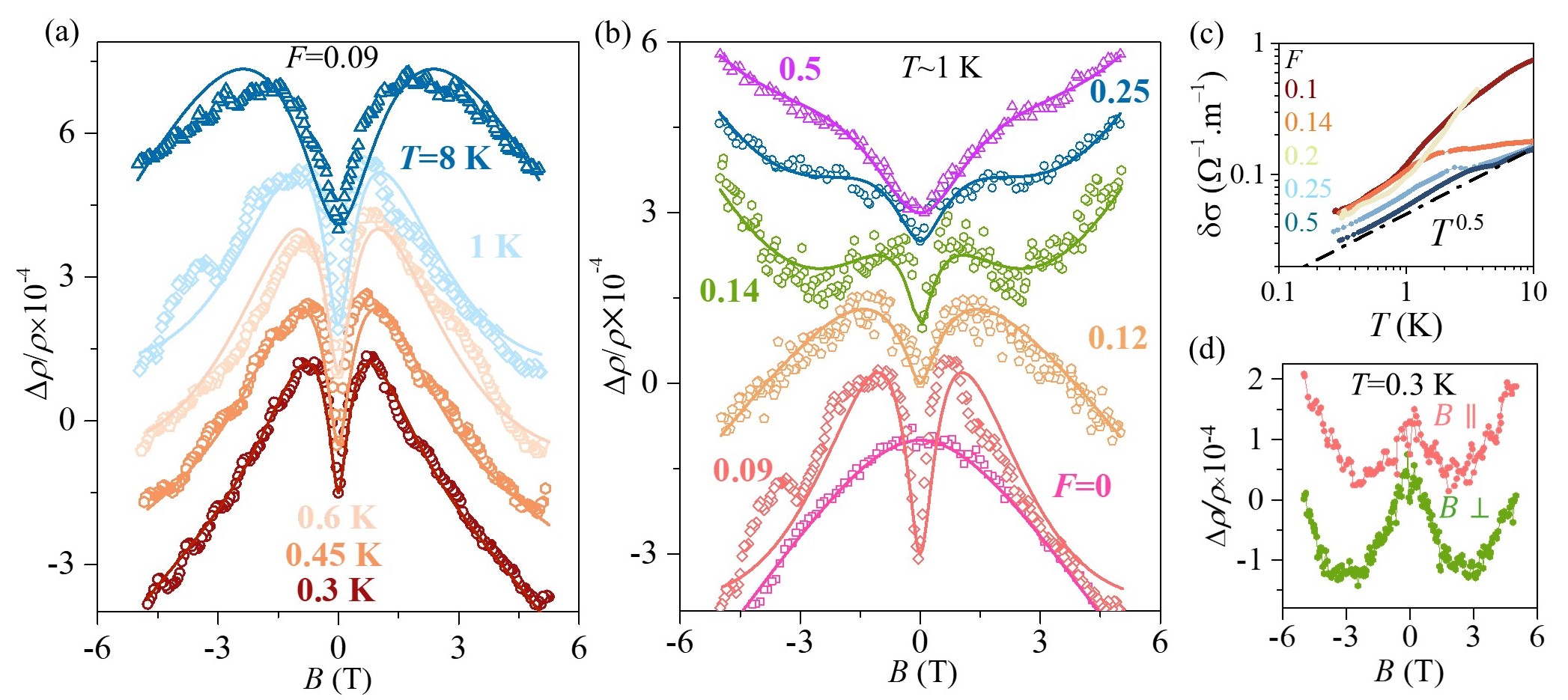}
    \caption{\textbf{Quantum transport}: (a) Magnetoresistance(MR), represented by the relative change in resistivity ($(\rho(B)-\rho(0))/\rho=\Delta\rho/\rho$) with magnetic field($B$), in a perpendicular magnetic field is measured for a film with Ag/Au interface density, $F=0.09$ at the various temperatures indicated. (b) MR at $T\sim 1$~K is shown for different films with values of $F$ varying from $0$ to $0.5$. $F=0$ represents a pure Au nanoparticle film. All the data in (a) and (b) have been shifted vertically for visual clarity. Solid lines in (a) and (b) represent fits to the data using conductivity corrections from quantum interference and electron-electron interaction effects(Eq.~[\ref{MR_total}]). (c) Low-$T$ transport for different values of $F$ is shown. This is quantified by the correction in conductivity, $\delta\sigma=-\delta\rho/\rho^2$, which is the change in conductivity ($\sigma=1/\rho$) with respect to $T$, from its value at $T=0$~K ($\sigma_0$), given by $\delta\sigma=\sigma-\sigma_0$. $\delta\rho$ indicates the change in $\rho$ with respect to $T$. $\sigma_0$ is estimated by extrapolation of the data to $T=0$. (d) MR, measured at $T=0.3$~K, is shown for a film with $F=0.14$ in magnetic fields applied parallel and perpendicular to the film.
    }
    \label{fig2}
\end{figure*}
 Fig.~\ref{fig1}\textcolor{blue}{(f)} (left axis) plots the $T$-dependent component of resistivity, $\rho_{300~\mathrm{K}}-\rho_{0}$,
 which represents the phonon contribution to resistivity with $F$. 
 To understand if the strong enhancement in $\rho_{300~\mathrm{K}}-\rho_{0}$ is due to an apparent enhancement in the electron-phonon coupling at intermediate values of $F$~\cite{kumbhakar2025engineering,mandal2024possibilitiesenhancedelectronphononinteractions,bose2024enormousenhancementresistivitynanostructured}, we simultaneously show the derivative of $\rho(T)$ with $T$ (taken at $T\sim\Theta_D=150K$, where $\Theta_D$ is the Debye temperature) on the right axis of Fig.~\ref{fig1}\textcolor{blue}{(f)}.
Since $\rho\approx2\pi m k_B \lambda T/\hbar n e^2$, where $n$ and $m$ are the electron's number density and mass, respectively, and $\lambda$ is the electron-phonon coupling (EPC) constant \cite{10.1093/acprof:oso/9780198507796.001.0001}, the slope of the $\rho-T$ curve directly corresponds to $\lambda$. The correspondence between $\rho_{300~\mathrm{K}}-\rho_{0}$ and $\mathrm{d}\rho/\mathrm{d}T$ suggests strong effective EPC, enhanced by nearly $\sim100$ times compared to that in bulk Ag or Au. Such an increase in EPC was recently suggested to arise from the charge transfer across Ag/Au interfaces~\cite{kumbhakar2025engineering,mandal2024possibilitiesenhancedelectronphononinteractions}.  
The impact of the interfaces and large EPC on the spin- and spin-orbit-related processes can be nontrivial, especially due to the breaking of structural inversion symmetry at the interfaces.

\subsubsection*{Magnetotransport measurements}
To probe the SOC, we have performed magnetotransport measurements down to $T\sim0.3$~K, with which we probe the phase-coherent processes and the spin-orbit scattering (See \textbf{SI} Section IV for details on the measurement). The magnetoresistance (MR) for a film, captured by $\Delta\rho/\rho$, with $F=0.09$ is shown in Fig.~\ref{fig2}\textcolor{blue}{(a)} at $T$ varying within $0.3-8$~K. At all $T$, we observe the MR increases with magnetic fields ($B$) at lower $B$ and then decreases at higher values of $B$. These represent the well-studied weak antilocalization (WAL) and weak localization (WL) behaviours observed in disordered conductors at low temperatures due to quantum interference effects \cite{bergmann1982inelastic,doi:10.1143/JPSJ.50.2516,BERGMANN19841,doi:10.1143/JPSJ.50.2131,baxter1989fitting}. (See \textbf{SI} Section V-VI for results on other values of $F$). 
The probability of the interference thus depends on the phase-breaking timescale, $\tau_{\phi}$. The nature of localization, on the other hand, \textit{i.e.} WL or WAL, is determined by the spin-orbit scattering time, $\tau_{\mathrm{soc}}$, which introduces a relative phase in the electronic wavefunctions.
Specifically, positive MR or WAL emerges from higher SO scattering \textit{i.e.} $\tau_{\mathrm{soc}}<\tau_{\phi}$ and vice-versa. Interestingly, we observe from Fig.~\ref{fig2}\textcolor{blue}{(a)} that the WL component decreases with increasing $T$, whereas the WAL component is almost constant. While the former behaviour of MR indicates decreasing $\tau_{\phi}$ with increasing $T$, which is generally observed across different systems, the latter dependence shows an unusual enhancement in the SOC with increasing $T$.

\begin{figure*}[ht]
    \centering
    \includegraphics[width=0.93\linewidth]{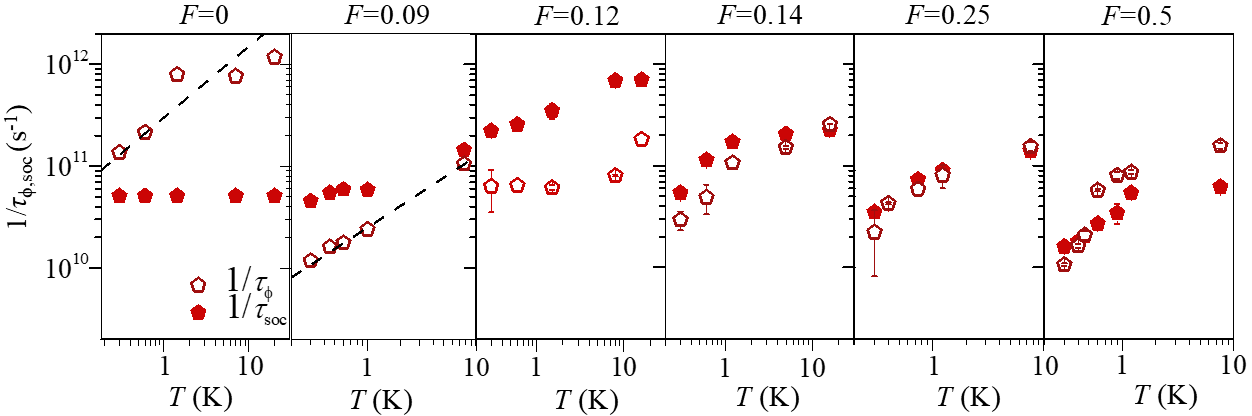}
    \caption{\textbf{Phase-breaking and Spin-orbit interaction timescales}: Temperature ($T$)-dependences of spin-orbit interaction rate ($1/\tau_{\mathrm{soc}}$) and phase-breaking rate ($1/\tau_{\phi}$) are shown for varying values of $F$. $\tau_{\mathrm{soc}}$ and $\tau_{\phi}$ are estimated from the fits to the quantum transport data, shown in Fig.~\ref{fig2}\textcolor{blue}{(a)} and \ref{fig2}\textcolor{blue}{(b)} using Eq.~[\ref{MR_total}]. 
    The black dashed lines in the first two panels for $F=0,0.09$ represent $T^{-0.7}$ dependence of $1/\tau_{\phi}$. 
    }
    \label{fig3}
    \hspace{0.3cm}
\end{figure*}

Fig.~\ref{fig2}\textcolor{blue}{(b)} shows the MR for films with different values of $F$ at $T\sim1$~K. For the pure Au nanoparticle (Au NP) film ($F=0$), we observe negative MR throughout the entire $B$ range, indicating WL. With increasing $F$, we find a distinct emergence of positive MR at low-$B$, suggesting WAL and, hence, enhanced SOC strength. Since Au, being heavier than Ag, has a higher intrinsic SOC \cite{browne2021quantum}, this increase of SOC with the incorporation of Ag in the hybrid system points toward an unconventional mechanism of spin-orbit scattering. We also observed another upturn in MR at higher magnetic fields for $F\gtrsim0.12$, which we attribute to the effects of the electron-electron interaction (EEI) \cite{RevModPhys.57.287,efros2012electron}. 
The proportionality of $\delta\sigma$ at low $T$ with $T^{1/2}$, shown in Fig.~\ref{fig2}\textcolor{blue}{(c)} also suggests strong EEI with increasing $F$~\cite{RevModPhys.57.287,10.1117/12.2063426}.

To quantitatively estimate $\tau_{\phi}$ and$\tau_{\mathrm{soc}}$, we have fitted the quantum transport data with additive conductivity corrections from WL/WAL and EEI effects \cite{PhysRevLett.108.106601,dey2014strong}. Since conductivity correction from EEI essentially arises from the Zeeman splitting of the spin levels, we have considered the Maekawa-Fukuyama form of WL/WAL \cite{doi:10.1143/JPSJ.50.2131}. Furthermore, we have considered a 3D model for both effects \cite{baxter1989fitting,efros2012electron,RevModPhys.57.287} since the MR in parallel and perpendicular fields are essentially the same, showing isotropic MR, as shown in Fig.~\ref{fig2}\textcolor{blue}{(d)}.
We fit the measured quantum correction to conductivity as follows.
\begin{equation}
    \frac{\Delta\rho}{\rho^2}=\frac{\Delta\rho}{\rho^2}\Big|_{\mathrm{QI}}+\frac{\Delta\rho}{\rho^2}\Big|_{\mathrm{EEI}}
    \label{MR_total}
\end{equation}
Here, $\frac{\Delta\rho}{\rho^2}\Big|_{\mathrm{QI}}$ and $\frac{\Delta\rho}{\rho^2}\Big|_{\mathrm{EEI}}$ are the conductivity corrections from quantum interference and EEI effects, respectively. The analytical forms of the fit equation can be found in the \textbf{Materials and Methods} section. It is to be noted that the quantum interference effects \textit{i.e} WL/WAL and the EEI effects differ in $T$ and $B$-dependences. While quantum interference gets suppressed with increasing $T$ (Fig.~2a and \textbf{SI} Section V) as the dephasing rate of the electron increases, the magnetic field-dependent component of EEI enhances with $T$ and always exhibits positive MR. Fits to the MR data using Eq.~[\ref{MR_total}] with four fitting parameters $B_{\phi}, B_{\mathrm{soc}},g,$ and $\mathcal{F}_{\sigma}$ are shown by the solid lines in Fig.~\ref{fig2}\textcolor{blue}{(a),(b)}, where $B_{\phi}=\hbar/4eD\tau_{\phi}$ and $B_{\mathrm{soc}}=\hbar/4eD\tau_{\mathrm{soc}}$ are the scales of phase-breaking magnetic field and spin-orbit magnetic field, respectively, $\mathcal{F}_\sigma$ is the average Coulomb interaction over the Fermi surface, and $g$ is the Lande $g$-factor. [See \textbf{SI} section V-VI for details on the fits and fitting parameters at all values of $F$ and $T$.] Diffusivity, $D=mv_F^2/3ne^2\rho$, that is dependent on $F$ has been estimated from the measured $\rho$, and constant number density, $n=10^{28}$~m$^{-3}$, electronic mass, $m=10^{-30}$~kg, and Fermi velocity of Au, $v_F=1.4\times10^6$~m.s$^{-1}$. 
(See \textbf{SI} Section IV for Hall measurements showing the estimation of $n$.)

\begin{figure}
    \centering
    \includegraphics[width=0.85\linewidth]{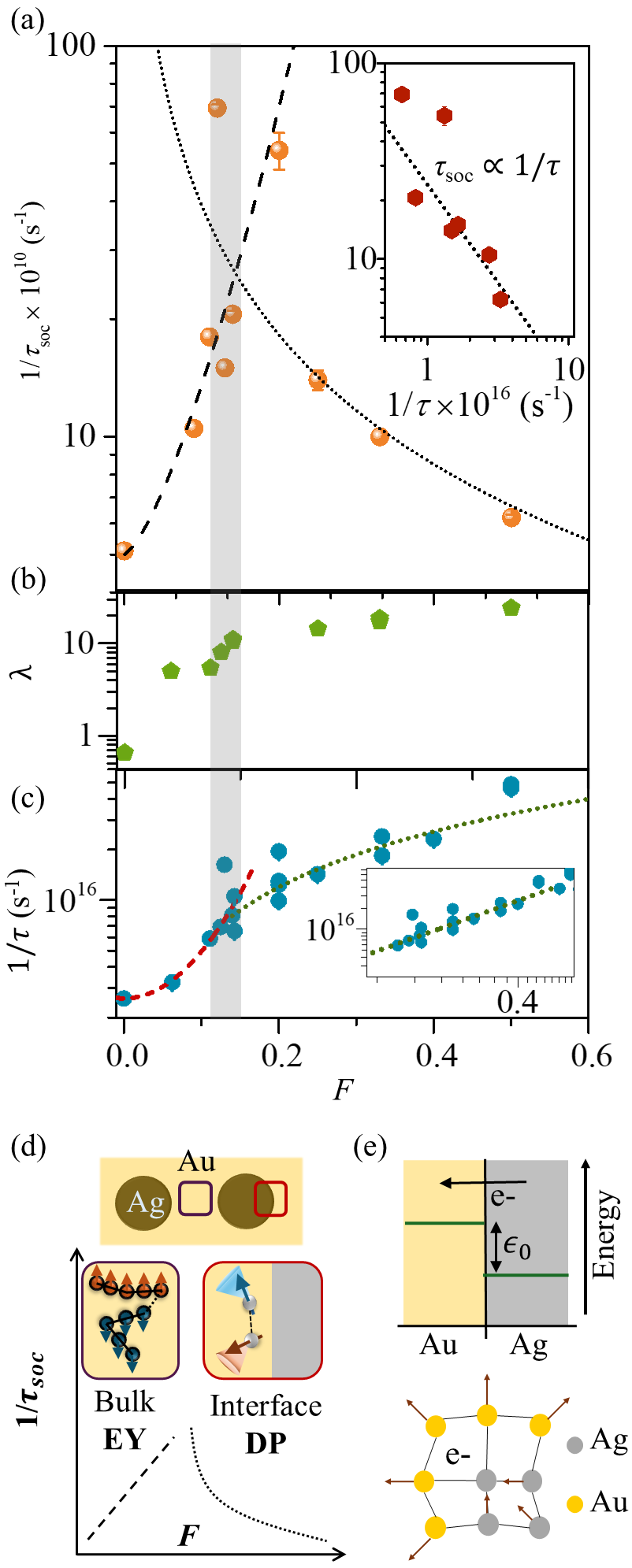}
    \label{fig4}
\end{figure}
\begin{figure}
    \centering
    \caption{\textbf{Rashba spin-orbit coupling:} (a) Top panel shows the variation of the spin-orbit scattering rate ($1/\tau_{\mathrm{soc}}$)
with $F$ at $T\sim8$~K. (b) shows electron-phonon coupling (EPC) constant ($\lambda$), estimated from $[\mathrm{d}\rho/\mathrm{d}T]_{150~\mathrm{K}}$ (Fig.~\ref{fig1}\textcolor{blue}{(f)}) at different values of $F$ using the expression $\rho_{T>\Theta_D}=2\pi \lambda mk_B T/ne^2\hbar$. (c) plots the electron scattering rate ($1/\tau$), estimated from the residual resistivity, $\rho_0$ (Fig.~\ref{fig1}\textcolor{blue}{(e)}) using the Drude expression of resistivity. The dashed and dotted lines in (a) represent exponential and linear dependences on $F$, respectively. The linear dependence of $1/\tau$ on $F$ is further illustrated in the inset of (c). The dashed and dotted lines in (a) represent $1/\tau_{\mathrm{soc}}\propto 1/\tau$ and $1/\tau_{\mathrm{soc}}\propto\tau$, respectively, with the expressions derived from the extrapolations of $1/\tau$ vs $F$ in (c). The inset in (a) shows the proportionality of $1/\tau_{\mathrm{soc}}$ with $\tau$ for $F\gtrsim0.11$.
$\alpha_R\approx15$~meV.\r{A} is estimated from the slope of the dotted line using the expression of DP spin relaxation,$\tau_{\mathrm{soc}}=3\hbar^4/8\alpha_R^2 m^2 v_F^2 \tau$. (d) Different mechanisms of spin relaxation are illustrated. At lower $F$, the EY mechanism originating from disorder potential in the bulk (purple box) dominates, whereas the DP mechanism from Rashba coupling at the interface (red box) is significant at higher-$F$. 
\textcolor{black}{(e) (Top panel) A schematic of the electrochemical potential of electrons at the Ag and Au sites across the Ag@Au nanohybrid is shown, $\epsilon_0$ being the potential difference between them. Electrons transfer from a higher onsite potential at Ag to a lower potential in Au. 
(Bottom panel) demonstrates the polaronic trapping of electrons at the Ag/Au interface. 
}
}
    \label{fig4}
     \hspace{0.1cm}
\end{figure}

Fig.~\ref{fig3} shows the $T$-dependence of $\tau_{\phi}$ and $\tau_{\mathrm{soc}}$ for different films with $F$ ranging within $0-0.5$. 
The $T$-dependence of $\tau_{\phi}$ follows $\sim T^{-p}$ behaviour with $p\sim0.7-1$ for lower Ag-filling ($F\lesssim0.09$), indicating that electron-electron interaction likely dominates the phase-breaking mechanism \cite{RevModPhys.57.287}. With increasing $F$, the 
$T$-dependence of $1/\tau_{\phi}$ deviates from a pure power law, becoming the weakest for $F\sim0.12-0.14$, which suggests strong modification of the phase-breaking processes.
(See \textbf{SI} Section VI for results at other values of $F$).
Secondly, the SO scattering rate, $1/\tau_{\mathrm{soc}}$, first increases with increasing F as compared to that of pristine Au, 
until $F\sim0.12$, after which it decreases again.
We find this behaviour at all temperatures, indicating it to be related to the nanostructuring of Au.
Finally, we find an emerging $T$-dependence in $1/\tau_{\mathrm{soc}}$ at larger $F$, which again points towards an unconventional origin of SOC in the system, likely due to the presence of interfaces.

Fig.~\ref{fig4}\textcolor{blue}{(a)} shows the $F$-dependence of $1/\tau_{\mathrm{soc}}$ (for $T\sim8$~K), where the nonmonotonicity of $1/\tau_{\mathrm{soc}}$ in $F$ is evident. $1/\tau_{\mathrm{soc}}$ first increases with increasing $F$, reaching $\sim7\times10^{11}$~s$^{-1}$ at $F=0.12$, which is more than an order of magnitude larger compared to that of pure Au NP film ($1/\tau_{\mathrm{soc}}\sim5\times10^{10}$~s$^{-1}$). Further increase in $F$ causes $1/\tau_{\mathrm{soc}}$ to decrease, approaching $\sim5.5\times10^{10}$~s$^{-1}$ at $F=0.5$. Intriguingly, we find that both $\lambda$, estimated from the slope of $\rho-T$ in Fig.~\ref{fig1}\textcolor{blue}{(f)}, and $1/\tau$ (calculated from the residual resistivity in Fig.~\ref{fig1}\textcolor{blue}{(e)}), undergo a change in their respective behaviour at very similar values of $F$. As shown in Fig.~\ref{fig4}\textcolor{blue}{(b)}, $\lambda$ increases sharply from $\sim0.6$ at $F=0$ to $\sim10$ at $F=0.14$, after which the increase in $\lambda$ becomes weaker. Similarly, $1/\tau$ also increases in a super-linear manner with $F$ until $F\sim0.13$ (red dashed line), after which it becomes almost linear (green dotted line, also in the double logarithmic plot in the inset of Fig.~\ref{fig4}\textcolor{blue}{c}). 
Since the linear increase in $1/\tau$ with $F$ has been quantitatively shown to correspond to interface-dominated scattering \cite{doi:10.1021/acsaelm.3c00379}, the nonmonotonic $F$-dependence of $1/\tau_{\mathrm{soc}}$ seems to represent a crossover in the nature of transport at $F\sim 0.13$, from bulk disorder dominated to that determined by the buried Ag/Au interfaces.

\begin{figure*}[ht]
    \centering
    \includegraphics[width=0.99\linewidth]{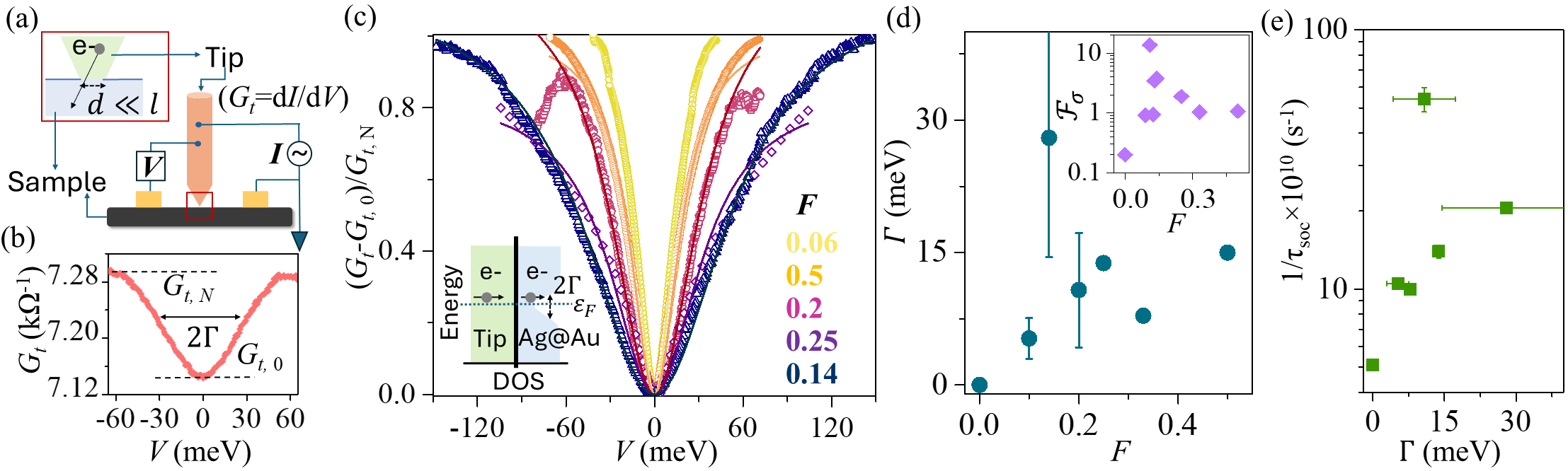}
    \caption{\textbf{Tunnelling measurements}: (a) The schematic of the experimental set-up. The tip is biased with a voltage ($V$) while the sample is grounded, thereby driving a current ($I$) across the tip-sample nanocontact of dimension $d$ and tunnelling conductance, $G_\mathrm{t}=\mathrm{d}I/\mathrm{d}V$. 
    (b) Typical $G_{{t}}$ for a films with $F = 0.14$, measured at $T\sim8$~K. (c) $G_{{t}}$ is shown for different films with varying $F$, measured at $T\sim8$~K. $G_{{t}}$ is normalized such that the minimum value at zero bias, $G_{t,\mathrm{0}}$ is $0$ and the value saturating at larger bias, $G_{t,\mathrm{N}}$ is $1$. 
    Solid lines are Lorentzian fits to the spectra with fitting parameter, $\Gamma$ representing the width of \textit{soft-gap} in the electronic DOS, illustrated schematically in the inset. 
    (d) Variation of $\Gamma$ with $F$ is shown. 
    Inset: $\mathcal{F}_{\sigma}$, representing the Coulomb screening parameter, estimated from quantum transport measurements. (e) Variation of $1/\tau_{\mathrm{soc}}$ at $T\sim8$~K with $\Gamma$ measured for different films with varying $F$. 
    }
    \label{fig5}
     \hspace{0.1cm}
\end{figure*}
Spin relaxation mechanisms are widely classified into two types (see the schematic of Fig.~\ref{fig4}\textcolor{blue}{(d)}). The first is the Elliot-Yafet (EY) mechanism \cite{PhysRev.96.266,YAFET19631}, where the conduction electron spin interacts with its motion in the electric field of the host lattice described by the periodic potential.
Here, spins of the conduction electrons relax/change due to momentum-scattering events that give rise to a spin-orbit scattering rate directly proportional to the momentum-scattering rate ($\tau_{\mathrm{soc}}\propto{\tau}$). The other is the Dyakonov-Perel (DP) mechanism \cite{DYAKONOV1971459,dyakonov1972spin}, where inversion symmetry breaking from Rashba coupling leads to a momentum-dependent effective magnetic field, which causes spin precession and resultant dephasing.
Momentum scattering, in this case, causes rapid fluctuations in the internal magnetic field and disruption of the spin dephasing.
Thus, $\tau_{\mathrm{soc}}$ induced by a Rashba interaction becomes longer
as the electron scattering time, $\tau$, becomes shorter ($\tau_{\mathrm{soc}}\propto{1/\tau}$).
The inset of Fig.~\ref{fig4}\textcolor{blue}{(a)} shows the dependence of $1/\tau_{\mathrm{soc}}$ on $1/\tau$, where $1/\tau$ is calculated from the magnitude of $\rho_0$, consistent with the fitting of the MR data, for films with $F > 0.12$. The inverse relation (dotted line) indicates the DP-type spin relaxation, which is consistent with the possibility of structural symmetry breaking at the Ag/Au interfaces. The dotted line in Fig.~\ref{fig4}\textcolor{blue}{(a)} represents the same behavior by mapping $1/\tau$ onto $F$ (green dotted lines in Fig.~\ref{fig4}\textcolor{blue}{(c)} and inset). Quantitatively, one expects 
$\tau_{\mathrm{soc}}=3\hbar^4/8\alpha_R^2 m^2 v_F^2 \tau$ in three dimensions, which yields $\alpha_R\approx15$~meV.\r{A} (dotted line in the inset of Fig.~\ref{fig4}\textcolor{blue}{(a)}). This is in close agreement with the Rashba interaction observed and computed at the planar interfaces created by depositing (few-layer)Ag on Au(111) substrate~\cite{PhysRevB.83.165401}. 
Finally, the dashed line describing the sharp increase in $1/\tau_{\mathrm{soc}}$ at low $F$ in Fig.~\ref{fig4}\textcolor{blue}{(a)} is representative of $\tau_{\mathrm{soc}} \propto \tau$, {\it i.e.} the EY mechanism, obtained by empirically fitting the $1/\tau$ data to $F$ at low values ($F \lesssim 0.12$, red dashed line in Fig.~\ref{fig4}\textcolor{blue}{(c)}).
\textcolor{black}{Fig.~\ref{fig4}\textcolor{blue}{(d)} depicts the two regimes of spin-relaxation observed at different regimes of $F$. Fig.~\ref{fig4}\textcolor{blue}{(e)} illustrates two key physical phenomena that could be intimately linked to the breaking of structural inversion symmetry and origin of the Rashba coupling. The top panel shows the charge transfer at the Ag/Au interface, driven by the difference of onsite electrochemical potential ($\epsilon_0$) (a theoretical estimation of the charge occupancy is shown in the \textbf{SI} section VII), and bottom panel demonstrates possible polaronic distortion at the interface, linked to the giant electron-phonon coupling reported in these structures~\cite{kumbhakar2025engineering}.}

Detailed studies have shown that large Rashba splitting is observed for surface states of noble metals like Au, Ag, and Cu \cite{PhysRevLett.77.3419,PhysRevB.63.115415,PhysRevB.65.033407,PhysRevB.87.075113,PETERSEN200049,BIHLMAYER20063888,PhysRevB.90.235422,Nagano_2009,PhysRevB.66.245419} that are truly localized at the surface, and their asymmetric nature across the surface crucially determines the magnitude of the Rashba effect. This can be understood in a simple way by relating the gradient of the wave function to the electric field close to the atomic core, which leads to the momentum-dependent magnetic field $\sim (\vec{k}\times\hat{z})$ \cite{PhysRevB.90.235422,PETERSEN200049,BIHLMAYER20063888,Nagano_2009}. In fact, confinement of the electron gas close to the interface has been found to enhance Rashba splitting in oxide and semiconductor heterostructures as well \cite{PhysRevLett.112.086802}.
Hence, the dominance of the Rashba effect with increasing interface density, $F$ of the Ag@Au hybrid structure, could accompany increasing localization of electronic wave functions close to the interface with higher $F$. 
\subsubsection*{Tunnelling measurements}
We have probed the electronic density of states (DOS) by performing tunnelling measurements on the films \cite{naidyuk2005point,kumbhakar2025engineering} to obtain evidence of such localization and its connection with the SOC. 
The experimental schematic is illustrated in Fig.~\ref{fig5}\textcolor{blue}{(a)}. A sharp metallic tip is brought in contact with the sample to measure the differential conductance ($G_t=\mathrm{d}I/\mathrm{d}V$) by passing a current, $I$, through the tip at a particular bias, $V$.  The dimension of the tip-sample contact was kept low for conductance to probe the tunnelling density of states (DOS). 
Fig.~\ref{fig5}\textcolor{blue}{(b)} shows the $G_t$ measured for a typical film of $F=0.14$ while varying $V$ within $\pm60$~meV at $T\sim8$~K. We observe a conductance minimum ($G_{t,0}$) at zero bias, which increases and saturates at larger bias, denoted by $G_{t,N}$. The full-width half maximum of the dip in conductance, represented by $\Gamma$ is $\approx30$~meV.
Fig.~\ref{fig5}\textcolor{blue}{(c)} shows the normalized tunelling spectra at $T\sim8$~K measured for varying  values of $F$. 
From the homogeneity of the dip across the film and its $T$-dependence (See \textbf{SI} Section VIII), we exclude possibilities of disorder/impurity-mediated tunelling for the conductance minima and attribute this to opening of a \textit{soft gap}, of width $\Gamma$, in the electronic DOS of the film, as represented schematically in the inset of Fig.~\ref{fig5}\textcolor{blue}{(c)}.
To extract $\Gamma$ precisely, we have fitted the spectra with a Lorentzian curve. 
Fig.~\ref{fig5}\textcolor{blue}{(d)} shows the estimation of $\Gamma$ at different values of $F$. We observe $\Gamma$ increases with increasing $F$, reaching a maximum of $\sim30$~meV close to $F\sim0.14$, after which it decreases again. 
The inset shows that the Coulomb screening parameter, $\mathcal{F}_{\sigma}$, estimated from quantum transport measurements, reaches a pronounced maximum close to $F\sim0.14$, where $\Gamma$ is also the highest. 
This further suggests that there is a depletion of itinerant electrons at the Fermi level, which results in a reduction of the Coulomb screening and enhanced $\mathcal{F}_{\sigma}$.


Polaronic localization of electrons close to the Ag/Au interfaces, like in small polarons, can enhance the wave function asymmetry, thus making Rashba coupling the dominant source of spin relaxation. Using both quantum transport and tunnelling measurements, we observe $\tau_{\mathrm{soc}}$ and $\Gamma$ to peak around $F\sim0.14$, implying the highest probability of localization and resulting interface-dominated electrical transport and inversion symmetry-breaking. This is quantitatively captured by the direct correspondence between $1/\tau_{\mathrm{soc}}$ and $\Gamma$ in Fig.~\ref{fig5}\textcolor{blue}{(e)}. It is possible that
the $T$-dependence of $1/\tau_{\mathrm{soc}}$ is also related to the temperature-driven dynamics of polarons, although further understanding is required. 

\subsection*{Discussion}
To summarize, we have demonstrated a novel metallic system where a Rashba SOC can be induced by a network of nanoscale interfaces. We show the coupling strength is tunable over an order of magnitude by varying the interface density ($F$), reaching a maximum value of $\sim15$~meV.\r{A}. 
We report, for the first time, Rashba coupling in a bulk metal that also globally preserves the inversion symmetry, though there is local inversion symmetry breaking by the buried interfaces. We have further demonstrated the crossover from EY-type to DP-type spin relaxation by tuning the interface density, which is usually difficult to achieve on the same material platform.
We show a unique interplay of electron-phonon coupling and Rashba interaction that can lead to many-body quantum states like spin-orbital polarons \cite{celiberti2024spin}. 
\subsection*{Materials and Methods}

\subsubsection*{Material synthesis:}

Ag@Au nanohybrids were synthesized via a two-step colloidal process\cite{doi:10.1021/acsaelm.3c00379,kumbhakar2025engineering}. First, AgNO$_3$ was reduced with ice-cold NaBH$_4$  in ultrapure water ($\sim$18.2 M$\Omega$·cm) containing NaOH, NH$_4$Br, KI, and CTAB to form AgNPs. Then,  HAuCl$_4$ was added at 40\textdegree C under stirring to form a gold shell. The reaction was monitored by UV–Vis spectroscopy and quenched with IPA, followed by centrifugation at 10,000 rpm for 30-45 minutes to remove excess CTAB.

The nanohybrids were drop-cast onto pre-patterned Cr/Au ($\sim$10/70 nm) electrodes on glass. Samples were dissolved in CHCl$_3$, dried at 60\textdegree C, and washed sequentially with DI water, KOH, and IPA to remove residual CTAB and promote sintering. This cycle was repeated ten times to form films of thickness, $t\approx3 \pm 0.5 \mu$m  (See Fig.~S2).

\subsubsection*{Electrical transport measurements:}

Four-probe resistivity measurements of the Ag@Au nanohybrids were carried out in home-built cryostats down to $\sim$ 6 K using a DC current of $\sim 100~\mu$A sourced by a Keithley 6221, while the voltage was measured using a Keithley 2182A nanovoltmeter. A Keithley 3700 multiplexer card enabled 
voltage acquisition across multiple contact pairs. Delta-mode voltage measurement was used to suppress thermal electromotive force (thermo-EMF)(See \textbf{SI} section IV). For lower temperatures ($T\sim$10 K to 0.3 K), resistivity and magnetotransport measurements were performed in a He-3 cryostat.
The tunelling measurements were performed in another home-built cryostat that can cool down to ~ 5 K and hosts a specialized tip-sample chamber with nanopositioners (attocubes and piezo
tubes) connected to the tip holder, and electrical contacts attached to both the tip and the film.
A sharp Pt/Rh metallic tip is
brought in contact with the film in a controlled manner with the help of the nanopositioners as indicated in the schematic of the experimental
set-up in Fig. 5(a) of main manuscript.

\subsubsection*{Analysis of quantum transport data:}

 Quantum transport data was analyzed using weak localization (WL), weak antilocalization (WAL), and electron-electron interaction (EEI) corrections to extract phase coherence time($\tau_\phi$) and spin-orbit scattering time($\tau_{\mathrm{soc}}$). The Maekawa-Fukuyama model was applied for WL/WAL effects, while a 3D model was used for both WL/WAL and EEI due to the isotropic magnetoresistance observed in the films. The conductivity corrections were then fitted to the experimental data.
The total quantum correction to conductivity is shown in Eq.~[\ref{MR_total}] of the main text.
Here, $\frac{\Delta\rho}{\rho^2}\Big|_{\mathrm{QI}}$ and $\frac{\Delta\rho}{\rho^2}\Big|_{\mathrm{EEI}}$ are the conductivity corrections from quantum interference and EEI effects, respectively~\cite{baxter1989fitting,doi:10.1143/JPSJ.50.2516}.
\\
\subsubsection*{1. Conductivity correction due to quantum interference:}

\begin{align}
  \frac{\Delta\rho}{\rho^2}\Big|_{\mathrm{QI}} 
  = &\; \frac{e^2}{2\pi^2\hbar} \sqrt{\frac{eB}{\hbar}} \Biggl[
    \frac{1}{2\sqrt{1 - \gamma}} \left( 
      f_3\left( \frac{B}{B_{-}} \right) 
      - f_3\left( \frac{B}{B_{+}} \right) 
    \right) \nonumber \\
  &\; - f_3\left( \frac{B}{B_2} \right) 
    - \sqrt{ \frac{4B_{\mathrm{soc}}}{3B} } \Biggl\{
    \frac{1}{\sqrt{1 - \gamma}} \left( 
      \sqrt{t_{+}} - \sqrt{t_{-}} 
    \right) \nonumber \\
  &\; + \left( 
      \sqrt{t} + \sqrt{t + 1} 
    \right) \Biggr\} 
  \Biggr]
  \label{MR_WL}
\end{align}

where 
\begin{equation}
   f_3 (z)=\Sigma_{n=0}^{\infty}\left[2(\sqrt{n+1+z}-\sqrt{n+z})-\frac{1}{n+\frac{1}{2}+z}\right]
\end{equation}

Here, $B_{\phi}=\hbar/4eD\tau_{\phi}$ and $B_{\mathrm{soc}}=\hbar/4eD\tau_{\mathrm{soc}}$ are the scales of phase-breaking magnetic field and spin-orbit magnetic field, respectively. $t=3B_{\phi}/4B_{\mathrm{soc}}$ qualitatively represents the strength of WL to WAL, $\gamma={(3g^{*}\mu_B B/8eDB_{\mathrm{soc}})}^2$ represents the strength of Zeeman splitting where $g^{*}$ is the effective g-factor. $B_{\pm}, B_2$ are characteristic fields given as: $B_{\pm}=B_{\phi}+\frac{2}{3}B_\mathrm{so}(1\pm\sqrt{1-\gamma})$, and $B_2=B_{\phi}+\frac{4}{3}B_{\mathrm{so}}$. It is to be noted that we have not considered any spin-flip scattering that can arise from magnetic impurities, thus implicitly making the phase-coherent and inelastic scattering processes equivalent \textit{i.e.} $\tau_i=\tau_{\phi}$, $\tau_{i}$ being the inelastic scattering time.
\\

\subsubsection*{2.Conductivity correction due to electron-electron interaction~\cite{RevModPhys.57.287}:}

\begin{align}
  \frac{\Delta\rho}{\rho^2}\Big|_{\mathrm{EEI}} = 
  \frac{8e^2}{3\pi^2\hbar} \mathcal{F}_\sigma 
  \Bigg( \left(1 + \frac{\mathcal{F}_\sigma}{2} \right)^{3/2} 
  & - 1 - \frac{3\mathcal{F}_\sigma}{4} \Bigg) \nonumber \\
  & \times \sqrt{\frac{T}{2D}} \, g_3(h)
  \label{MR_EEI}
\end{align}

where
\begin{equation}
    \mathcal{F}_\sigma=\frac{\int d \hat{\Omega} v (q=2k_F \mathrm{sin}(\theta/2))}{\int d\hat{\Omega} v(0)}
\end{equation}
is the average of the Coulomb interaction $v(q)$ on the Fermi surface over the solid angle $\hat{\Omega}$.

{$^\dagger$\textbf{Equal contribution}}\\
 {$^\ast$\textbf{Corresponding authors}:\\
shreyak@iisc.ac.in\\
banashreed@iisc.ac.in\\
arindam@iisc.ac.in}

\newpage
\bibliography{bibliography}

\section*{Author contributions}
S.K. and B.D. contributed equally to this work. 

S.K., B.D., and A.G. conceived the idea. S.K., T.K.M., B.D., and A.G. designed and carried out the magnetotransport experiments with help from T.P.S. . S.K., T.P.S., and A.G. designed and performed the tunelling experiments. T.K.M., S.K., and B.D. fabricated and characterized the samples. B.T performed the Transmission Microscopy measurements. S.M., M.J., and H.R.K. performed theoretical modelling. T.V.R. provided theoretical inputs. A.P. provided experimental inputs to the sample fabrication and contributed during the revision of the manuscript. S.K. and B.D. analysed the data.  S.K., B.D., and A.G. wrote the manuscript with inputs from all the authors.
\section*{Funding}
S.K. acknowledges support from the Prime Minister's Research Fellowship (PMRF). T.K.M.  acknowledges the National Postdoctoral Fellowship (NPDF) for funding. B.T. acknowledges the Department of Science and Technology, Government of India, through the INSPIRE Faculty Fellowship (DST/INSPIRE/04/002484). S.M., M.J., and H.R.K. thank the Supercomputer Education and Research Centre (SERC) at the Indian Institute of Science (IISc) for providing
the computational resources. M.J. acknowledges the National Supercomputing Mission of the Department of Science and Technology, India, and Nano Mission of the Department of Science and Technology for financial support under Grants No. DST/NSM/R\&D\_HPC\_Applications$/2021/23$ and No. DST/NM/TUE/QM-10/2019 respectively.  H. R. K. gratefully acknowledges support from the Indian National Science Academy under grant number No. INSA/SP/SS/2023/,
the Science and Engineering Research Board of the Department of Science and Technology, India, under grant No.
SB/DF/005/2017/, and at the International Centre for Theoretical Sciences (ICTS) from the Simons Foundation
(Grant No. 677895, R.G.). A.P. acknowledges the IISc Institute of Eminence (IOE) for funding and support.
\end{document}


\setcounter{figure}{0}
\renewcommand{\figurename}{Fig.}
\renewcommand{\thefigure}{S\arabic{figure}}
\renewcommand{\theequation}{S\arabic{equation}}
\renewcommand{\thepage}{S\arabic{page}}
\renewcommand{\tablename}{Table S}
\preprint{APS/123-QED}
\title{
{\Large Supplementary Information}\\
Emergent Rashba spin-orbit coupling in bulk gold with buried network of nanoscale interfaces}
\author{Shreya Kumbhakar$^{1,\dagger, \ast}$}
\author{Banashree Debnath$^{1,\dagger, \ast}$}
\author{Tuhin Kumar Maji$^{1}$}%
  \author{Binita Tongbram$^1$}%
   \author{Shinjan Mandal$^1$}
  \author{T. Phanindra Sai$^1$}%
  \author{T.V. Ramakrishnan$^1$}
 \author{Manish Jain$^1$}
 \author{H. R. Krishnamurthy$^{1,2}$}
  \author{Anshu Pandey$^3$}%
 \author{Arindam Ghosh$^{1,\ast}$}%
\affiliation{%
 $^1$Department of Physics, Indian Institute of Science, Bangalore 560012, India
}%
\affiliation{%
 $^3$Solid State and Structural Chemistry Unit, Indian Institute of Science, Bangalore 560012, India
}
\affiliation{$^2$International Centre for Theoretical Sciences, Tata Institute of Fundamental Research, Bangalore 560089, India}
\affiliation{$^\dagger$\textbf{
authors contributed equally}}
\affiliation{}
\affiliation{$^\ast$\textbf{Corresponding authors}:
\\
  shreyak@iisc.ac.in\\
  banashreed@iisc.ac.in\\
  arindam@iisc.ac.in}
\maketitle
\section{Material Synthesis}
\noindent\textbf{Synthesis of AgNP:} 

Silver nanoparticles (AgNPs) were prepared by reducing AgNO$_3$ ($1$~mM) using ice-cold NaBH$_4$ ($0.1$~M) as a reducing agent. CTAB ($0.1$~M) was added as a stabilizing agent to control the nanoparticle size and uniformity. The reaction mixture, which also contained NH$_4$Br ($1$~M), was maintained at $40\degree$C, with its pH slightly adjusted to basic conditions using NaOH ($0.1$~M). To ensure a high-yield synthesis of uniform AgNPs, KI ($0.1$~M) was incorporated. Upon the introduction of AgNO$_3$, the solution turned whitish and turbid, signifying the formation of insoluble silver-halide (AgX) clusters. After a carefully monitored waiting period~\cite{doi:10.1021/acsaelm.3c00379}, NaBH$_4$ was gradually added, resulting in an instantaneous color shift to yellow, indicating AgNP formation. This was further verified by monitoring the surface plasmon resonance (SPR) of the colloidal solution, which has a distinct peak at $393$~nm from AgNPs~\cite{doi:10.1021/acsaelm.3c00379,kumbhakar2025engineering}. 
While preparing pure AgNP film, the reaction is terminated with isopropyl alcohol (IPA) after the formation of AgNPs. The solution was centrifuged at $10,000$~rpm for $60$~minutes to precipitate the nanoparticles.

\noindent\textbf{Synthesis of Ag@Au nanohybrid:}

To fabricate Ag@Au nanohybrids, HAuCl$_4$ ($1$~mM) was introduced into the AgNP solution, leading to a colour transformation from yellow to brown, confirming the formation of an Au shell around the Ag core. A red shift in the UV-Vis absorption spectrum further validated the formation of the Au shell and the formation of the Ag@Au nanohybrid (NH) structure~\cite{doi:10.1021/acsaelm.3c00379,kumbhakar2025engineering}. The reaction was terminated by adding isopropyl alcohol (IPA), facilitating nanoparticle precipitation. The solution was then subjected to centrifugation at $10,000$~rpm for $30-45$~minutes, after which the supernatant was removed, and the nanohybrids were collected for further processing.

\noindent\textbf{Synthesis of AuNP:}

The Au nanoparticles (Au NPs) were synthesized following the same protocol as AgNPs, with HAuCl$_4$ replacing AgNO$_3$ while keeping the other experimental parameters unchanged. The reduction of HAuCl$_4$ with NaBH$_4$ ($0.1$~M) induced a reddish-purple colour shift, marking the successful formation of Au NPs. The reaction was terminated by the addition of isopropyl alcohol (IPA), and the nanoparticles were precipitated by centrifugation at $10,000$~rpm for $60$~minutes. The obtained solid sample was then used for further processing~\cite {doi:10.1021/acsaelm.3c00379,kumbhakar2025engineering}.


\section{{Structural Characterization}}
\begin{figure}[hbt!]
    \centering
    \includegraphics[width=0.9\textwidth]{Fig_S1.png}
    \caption{High-Angle Annular Dark-Field imaging (HAADF) images of Ag@Au nanohybrids are shown at various values of \textit{F}: (a) $F=0.57$,(b) $F=0.2$, and (c) $F=0.13$.The darker crystalline region of Au hosts a dispersion of the brighter region of Ag of diameter $\sim2$~nm. We clearly observe the integrity of the spherical AgNPs in the embedded structure at all $F$. We also note the gradual decrease of the average separation between the embedded interfaces with increasing $F$, again pointing towards structural integrity with dense packing of interfaces.}
    \label{Fig_S1}
    \hspace{0.1cm}
\end{figure}

\section{Film preparation}
The Ag@Au NH films were prepared using a drop-casting method onto pre-patterned Cr/Au contacts($10$~nm/$70$~nm) in a Van der Pauw or Hall bar configuration on a glass substrate.
The substrate was prepared using optical photolithography, followed by metal deposition and a lift-off process to pattern the Cr/Au contacts. 
Before drop-casting, the Ag@Au NH solution was centrifuged, dried under vacuum ($\approx$1 mbar), redispersed in chloroform, and sonicated. The drop-casting process involved sequential drying at $50-60\degree$C, washing with DI water and KOH to remove CTAB, followed by rinsing in IPA~\cite{doi:10.1021/acsaelm.3c00379,kumbhakar2025engineering}.
This cycle was repeated  $5-6$~times, resulting in films with an average thickness of $\approx2-3\pm 0.5~\micro$m over a $1$~mm$\times$100~$\micro$m  area, as shown in Fig~\ref{Fig_S2}. We have scanned the thickness profile across the black double arrow line as shown in the inset of Fig~\ref{Fig_S2}. For the preparation of patterned nanoparticle films, Ag@Au NHs were drop-cast onto a patterned photoresist-coated substrate, followed by multiple cleaning cycles ($5–6$~times) as mentioned to remove excess residual chemicals. Finally, the polymer was removed using thermal treatment, leaving behind the patterned nanoparticle film on the Cr/Au leads. Finally, all films were coated with PMMA and sealed with a glass cover slip for encapsulation.

\begin{figure}[hbt!]
    \centering
    \includegraphics[width=0.4\textwidth]{Fig_S2.png}
    \caption{Line scan showing the thickness ($t$) profile of a typical film, shown in the inset, is measured using optical profilometry along the double-arrowed line.}
    \label{Fig_S2}
    \hspace{0.1cm}
\end{figure}

\section{Details of Electrical Transport Measurements}
\subsection{Electrical resistivity measurements}
Four-probe electrical transport measurements were performed by applying a DC current of $100$~$\mu$A using a Keithley 6221 source meter and measuring the voltage drop with a Keithley 2182A nanovoltmeter. A Keithley 3700 multiplexer card was used to record resistance across multiple channels simultaneously. Measurements were conducted in delta mode to minimize thermal electromotive force (EMF) errors, which cancel unwanted voltage offsets by alternating the current direction. For measuring the temperature dependence of the resistivity from 300~K to low temperatures. The experiment was carried out in a home-built variable temperature cryostat, capable of cooling the sample down to $\approx6$~K.
\subsection{Magnetotransport measurements}
Magnetotransport measurements were conducted in a He-3 cryostat with magnetic fields up to $\pm5$~T and temperatures ranging from $0.28–30$~K. An AC measurement technique was implemented where an SR830 lock-in amplifier was used to drive current through sample contact $I+$ at an excitation frequency of $11$~Hz. Notch filters were implemented to eliminate noise from line frequency ($\sim50$~Hz) and its second harmonic. At the same time, the voltage signal was amplified using an SR560 low-noise voltage amplifier with a gain factor of 500. A $1$~kHz low-pass filter (12 dB/octave roll-off) was implemented to improve the signal-to-noise ratio. The measurement schematic is shown in Fig.~\ref{Fig_S3}.
\begin{figure*}
    \centering
    \includegraphics[width=0.8\textwidth]{Fig_S3.png}
    \caption{Schematic of experimental set-up for magnetotransport measurements.}
    \label{Fig_S3}
\end{figure*}
The amplifiers were placed inside a Faraday cage for electromagnetic shielding. The Faraday cage is constructed of adjacent aluminium and mild steel sheets, both having a thickness of $1.5$~mm. The shielding effectiveness is determined by the skin depth ($\delta$)  of the particular metal at different frequencies. The thickness of the shield is typically kept as $3-4$ times $\delta$, ensuring effective shielding, where $\delta$=$\sqrt{1/\pi\sigma\mu f}$ ($f$ is the frequency, $\mu=\mu_0\mu_r$ is the permeability, and $\sigma$ is the conductivity of the metal). 
Using this, one obtains  $\delta_{\mathrm{Al}} \approx0.83$~mm at 10~kHz and $\delta_{\mathrm{steel}} \approx0.20$~mm at 1~kHz. Hence, the Faraday cage effectively screens out high-frequency noise. 

 For performing magnetotransport measurements, a strategy of \textit{hold and ramp} has been implemented to record the data. Specifically, the magnet is first ramped to the desired magnetic field, and the persistent heater is switched off. The data was recorded after a particular wait time (determined by the data acquisition rate) of \textit{holding} the magnet in persistent mode, ensuring a steady magnetic field and avoiding the possibility of inducing eddy currents. A lock-in time constant of $10$~s is used to average the signal, effectively filtering out high-frequency noise and improving the signal-to-noise ratio. The persistent switch heater is then switched on again, and the magnet is \textit{ramped} to the next desired field.
\subsection{Hall measurements}
\begin{figure*}[ht]
    \centering
    \includegraphics[width=0.4\textwidth]{Fig_S4.png}
    \caption{
    Hall measurements at $T\sim0.3$~K are shown for films with varying Ag volume fraction, $F=V_{\mathrm{Ag}}/(V_{\mathrm{Ag}}+V_{\mathrm{Au}})$. The inset shows the schematic of the lead geometry used for Hall measurements. Yellow lines show the deposited Cr/Au electrodes, and the light grey region is the dropcast film. $I+,I-$ and $V+,V-$ indicate the current and voltage probes respectively. The electron density is estimated to be $n~(\sim 10^{28}~\mathrm{m}^{-3})$ for all $F$-values from the measured Hall resistance ($V_\mathrm{H}/I=1/ne$, $V_{\mathrm{H}}$, and $I$ are the measured hall voltage and current, respectively).
.
    }
    \label{Fig_S4}
\end{figure*}
\section{Evidence of Quantum Transport}

Quantum transport refers to the manifestation of quantum interference effects in the electrical transport properties. Briefly, in disordered conductors, electrons can travel in self-intersecting loops due to enhanced momentum scattering by disorder. At low values of $T$, when the inelastic scattering rate or the dephasing rate of the electrons reduces, the electrons in the forward and backwards-travelling paths can interfere constructively or destructively, giving rise to WL or WAL, respectively. In materials with strong spin-orbit coupling, WAL dominates at low magnetic fields, while WL emerges at higher fields. Both temperature dependence and angle dependence are powerful tools to confirm the quantum nature of WL/WAL effects. Our temperature-dependent measurements, in Fig.~2a and Fig.~\ref{Fig_S5}, reveal that the WL contribution weakens with increasing temperature, consistent with reduced $\tau_{\phi}$ with decreasing $T$, observed in diverse materials, while the WAL component remains largely unaffected. Also, we note that at higher values of $F$, there is an enhancement of MR at higher magnetic fields. We attribute this to the electron-electron interaction effects (EEI), which are in competition with the quantum interference effects. It should be noted that the quantum interference effects \textit{i.e} WL/WAL and the EEI effects differ in $T$ and $B$ dependences [See Methods of the main manuscript]. While quantum interference gets suppressed with increasing $T$ (Fig.~2a and \textbf{SI} Section V) as the dephasing rate of the electron increases, the magnetic field-dependent component of EEI enhances with $T$ and always exhibits positive MR. Thus, the magnitude of MR alone is not a direct measure of WL/WAL. 
To make this point clearer, we present below in Fig.~\ref{Fig_S5}, the temperature dependence of MR up to $T\sim30$~K for a sample with $F=0.09$. Since EEI, which is not a quantum interference effect, is minimal here, the suppression of WL/WAL effects is clearer in this sample.

\subsection{Temperature dependence of MR}
\begin{figure}[ht]
    \centering
    \includegraphics[width=0.5\linewidth]{Fig_S5.png}
    \caption{Temperature dependence of MR of a film with Ag-filling fraction, $F=0.09$ from $T\sim0.3$~K to 30~K. We clearly observe the suppression of MR with increasing $T$.}
    \label{Fig_S5}
\end{figure}
\newpage
\subsection{Angle dependence of MR}
Angle-dependent MR measurements in Fig.~\ref{Fig_S6} show that the MR remains nearly unchanged between $\theta$= $0\degree$ (magnetic field parallel to the sample plane) and $90\degree$ (magnetic field perpendicular to the sample plane) for all temperatures, indicating isotropic behavior. This is not unexpected because of the three-dimensional nature of our film, where the thickness of the film exceeds the phase-breaking and other length scales by two to three orders of magnitude. This justifies the use of a 3D theoretical model, as discussed in the Methods of the main manuscript.
\begin{figure}[ht]
    \centering
    \includegraphics[width=0.8\linewidth]{Fig_S6.png}
    \caption{Angle-dependent MR of $F=0.14$ showing isotropic behavior at $T=0.3$~K.}
    \label{Fig_S6}
\end{figure}

\newpage
\section{Analysis of the fitting parameters of the magnetoresistance data}
$H_{\phi} (l_{\phi}),H_{\mathrm{soc}} (l_{\mathrm{soc}}), \mathcal{F}_{\sigma}$, and $g$ are the parameters obtained by fitting the magnetoresistance data shown in Fig.~\ref{Fig_S7} using Eq.~[1] of the main manuscript.
\begin{figure}[ht]
    \centering
    \includegraphics[width=1\linewidth]{Fig_S7.png}
    \caption{Magnetoresistance (MR), represented by the relative change in resistivity ($(\rho(B)-\rho(0))/\rho=\Delta\rho/\rho$) with magnetic field, in a perpendicular magnetic field is measured for films with varying Ag/Au interface density, $F=0,0.09,0.11,0.12,0.13,0.14,0.2,0.25,0.33,0.5$. $F=0$ represents pure AuNP film. Solid lines represent fits to the data using conductivity corrections from quantum interference and electron-electron interaction effects (Eq.~[1] of the main manuscript). The data in all panels have been shifted vertically for visual clarity.}
    \label{Fig_S7}
\end{figure}
\subsection{Phase-coherent and spin-orbit scattering rates}
\begin{figure}[ht]
    \centering
    \includegraphics[width=1\linewidth]{Fig_S8.png}
    \caption{Temperature ($T$)-dependences of spin-orbit interaction rate ($1/\tau_{\mathrm{soc}}$) and phase-breaking rate ($1/\tau_{\phi}$) are shown for varying values of  $F$. The dotted lines in all panels represent $1/\tau_{\phi}^{-p}$ with $p\sim0.7-1$.}
    \label{Fig_S8}
\end{figure}
\subsection{Phase breaking length}
The phase coherence length $l_{\phi}$ is estimated from the phase breaking magnetic field $H_\phi$ as $\l_\phi=\sqrt{\hbar/4eH_\phi}$.
For all values of $F$, $l_{\phi}$ ranges within $10-100$~nm with a consistent decrease towards higher-$F$, indicating a higher phase-breaking rate. 
\begin{figure*}[hbt!]
    \centering
    \includegraphics[width=0.5\textwidth]{Fig_S9.png}
    \caption{Temperature ($T$) dependence of the phase coherence length $l_{\phi}$ is shown for different values of $F$.}
    \label{Fig_S9}
\end{figure*}

\subsection{\large{Average screened Coulomb potential, $\mathcal{F}_\sigma$}}

\begin{figure*}[hbt!]
    \centering
    \includegraphics[width=0.5\textwidth]{Fig_S10.png}
    \caption{$\mathcal{F}_{\sigma}$, computed from fits to the MR data, is shown for different films with varying $F$.}
    \label{Fig_S10}
\end{figure*}
\newpage
\subsection{\large{g-values}}
The effective g-factor in a system is such that the energy between the Zeeman split spin states is $g_{\mathrm{eff}}=mu_B B$, $B$ being the applied magnetic field. For a free electron, $g=2$. However, in a solid-state system, this number can be drastically different. For example, if there is no orbital moment or if it is quenched, the spin-orbit interaction mixes the spin-up and spin-down states, thus reducing the g-factor \cite{PhysRev.96.266}. In the presence of orbital contributions, SOC increases the electron moment, thus increasing the values of $g$~\cite{doi:10.1021/nl901333a}. Additional factors contribute to low-dimensional systems as compared to bulk systems. For example, in metal nanoparticles, interface scattering reduces the orbital contributions, thereby decreasing $g$ \cite{PhysRevLett.85.2789,PhysRevLett.87.266801,PhysRevLett.83.1644}. Such quantum confinement effects also give rise to mesoscopic fluctuations in the values of $g$~\cite{PhysRevLett.85.369}. It was observed that in Au nanoparticles the $g$-factor is reduced from the free electron value of $2$ to about $0.3$ \cite{PhysRevLett.83.1644,PhysRevLett.87.266801}. All the discussions above considered electron-electron interactions. In the absence of orbital contribution, the effects of EEI and SOC are competing in nature as the former tends to lift the spin degeneracy via exchange interactions thus increasing the net spin of a many-electron state and hence the effective $g$-value, whereas the latter mixes the spin states, thus decreasing $g$ \cite{PhysRevB.69.155417}.\\
In our experiments, we observe $g$ for Au nanoparticle film to be $\sim 0.2$, consistent with previous studies on Au nanoparticles \cite{PhysRevLett.83.1644,PhysRevLett.87.266801}. With finite Ag filling at $F=0.09$, the $g$-factor increases and has an increasing $T$-dependence as well. Interestingly, we also observe $g$ to increase till $F\sim0.14$, and it decreases again for $F=0.5$. This resembles the dependence of $\tau_{\mathrm{soc}}$ on $F$ and is consistent with a spin-orbit coupled state where the orbital contribution and strong SOC increases values of $g$ \cite{doi:10.1021/nl901333a}.
\begin{figure*}[hbt!]
    \centering
    \includegraphics[width=0.5\textwidth]{Fig_S11.png}
    \caption{Temperature ($T$)-dependence of the effective electron Lande g-factor is shown for films with varying $F$.}
    \label{Fig_S11}
\end{figure*}
\section{Details of theoretical computation}
\begin{figure}
    \centering
    \includegraphics[width=0.5\linewidth]{Fig_S12.png}
    \caption{Excess electron occupancy $\delta \mathrm{\textbf{n}_i}$ at each atomic site $i$, of a 2D square lattice with Ag embedded within a matrix of Au atoms.}
    \label{Fig_S12}
\end{figure}
To estimate the charge transfer at the Ag/Au interfaces due to the difference of onsite electrochemical potentials, the electronic structure was modeled using a semi-phenomenological tight-binding model
constructed from the s-like conduction bands of Au and Ag. The model includes uniform nearest neighbor hopping, site energy offsets to account for the work function mismatch between the two
elements, and both on-site and long-range Coulomb interactions. These interactions were treated
within the Hartree approximation, enabling a self-consistent determination of the Hartree potentials
and the resulting charge redistribution across the lattice sites.
To facilitate analytical exploration and for the ease of studying the effects of parameter variation, we
studied a simplified two-dimensional square superlattice, consisting of periodic (4×4) Ag clusters
embedded within an Au matrix, forming a 64-site unit cell. This toy model captures the essential physics of interface-induced charge inhomogeneity and enables controlled studies of electron-electron and electron-phonon coupling effects. The charge transfer in this lattice for an onsite electrochemical potential difference, $\epsilon_0=1.4$~eV, is shown in Fig.~\ref{Fig_S12}. 
From the excess electron occupancy ($\delta\mathrm{\textbf{r}_i}$) profile, we note electron doping from Ag to Au, as depicted schematically in Fig.~4 of the main manuscript.
For more details on the theoretical calculations, please refer
to \cite{mandal2024possibilitiesenhancedelectronphononinteractions}.
\newpage
\section{Tunelling Measurements}
The tunelling measurements were performed by bringing a sharp Pt/Rh metallic tip close to the film in a controlled manner with the help of nanopositioners (attocubes and piezoceramic cylinders) as indicated in the schematic of the experimental set-up in Fig.~5(a) of the main manuscript. The tip contacts the sample with an effective diameter $d$. The tip-sample chamber is loaded inside a home-built cryostat that could be cooled down to $T\sim5$~K. 
We have measured the differential resistance/conductance across the tip-sample contact in a four-probe configuration with the modulation spectroscopy technique \cite{naidyuk2005point,kumbhakar2025engineering}. Specifically, a mixed AC+DC current, $I+\delta I cos (\omega t)$, where $I$ is the DC current, $\delta I$ is the AC current and $\omega\sim227$~Hz is the AC excitation frequency is sent to the tip while the sample is grounded. The AC bias, $\delta V$ across the tip-sample contact, is measured in a four-probe configuration at $\omega$ frequency and $0^{\circ}$ phase under the corresponding DC biasing, giving the differential resistance as $R_{t}=\delta V/\delta I$.
The DC current is varied to tune the DC bias ($V$) across the tip-sample contact in the desired energy ($eV$) range. A typical spectrum measured for film $F=0.14$ at $T\sim8$~K is shown in Fig.~\ref{Fig_S13}\textcolor{blue}{(a)}. 
\begin{figure}[ht]
    \centering
    \includegraphics[width=0.9\linewidth]{Fig_S13.PNG}
    \caption{(a) The tunelling $I-V$ characteristics and the corresponding bias dependence of measured tunelling resistance, $R_t=\mathrm{d}V/\mathrm{d}I$ are shown for a typical film with Ag volume fraction $F=0.14$ at $T\sim8$~K. (b) shows the bias dependence of normalized $R_t$ \textit{i.e.} $(R_t-R_{t,N})/R_{t,0}$ measured at distinct positions of the sample, that collapse on each other.}
    \label{Fig_S13}
   \hspace{0.1cm}
\end{figure}
Fig.~\ref{Fig_S13}\textcolor{blue}{(b)} shows the tunelling spectra ($R_t$ vs $V$) measured at different positions in the sample. $R_t$ is normalized such that the resistance maximum ($R_{t,0}$) at zero bias is $1$ and the value ($R_{t,N}$) saturating at higher bias is $0$. The collapse of the gap measured at different positions excludes any disorder-mediated scattering. $R_t$ is inverted to obtain the tunelling conductance as $G_t=1/R_t$.

\begin{figure}[ht]
    \centering
    \includegraphics[width=0.43\linewidth]{Fig_S14.PNG}
    \caption{Comparison of elastic ($l_{\mathrm{el}}$), inelastic ($l_{\mathrm{in}}$) length scales, and the tip-sample contact diameter ($d$) for different values of Ag-filling, $F$ at temperature $T\sim8$~K. For $F\gtrsim 0.3, l_{\mathrm{el}}$ becomes less than the interatomic spacing, indicating the inapplicability of the Drude expression at these resistivities to estimate the elastic scattering length.}
    \label{Fig_S14}
    \hspace{0.1cm}
\end{figure}
Fig.~\ref{Fig_S14} shows a comparison between the elastic ($l_{\mathrm{el}}$), inelastic ($l_{\mathrm{in}}$) length scales with the tip-sample contact diameter, $d$ at different values. $d$ is estimated from a diffusive transport regime as $\rho/R_{t}$. $l_{\mathrm{in}}$ is equivalent to the phase-breaking length ($l_{\phi}$) that has been derived from the quantum transport measurements. We note that this equivalence holds under the assumption that there are no magnetic impurities in the system. {$l_{\mathrm{el}}=\sqrt{D\tau}$, where the diffusivity $D$ and scattering time $\tau$ are estimated from the residual resistivity via Drude expression as discussed in the main text. We observe $d<l_{\mathrm{in}}$, indicating that the collisions at the tip-sample contact are energy-conserving. Hence, the conductance directly maps the tunelling density of states, as illustrated schematically in the inset of Fig.~5(c) of the main manuscript.

Fig.~\ref{Fig_S15} shows the $T$-dependence of the tunelling spectra. The behaviour changes to that of a metallic differential resistance. This can be attributed to increased inelastic collisions and resultant energy-relaxing processes at the tip-sample contact, due to which the tunelling DOS cannot be probed.
The $T$-dependence of the spectra enables us to exclude any defect-mediated tunnelling process as the origin of the dip in tunelling conductance.
\begin{figure}[h!]
    \centering
    \includegraphics[width=0.5\linewidth]{Fig_S15.jpg}
    \caption{Tunnelling resistance ($R_t$) is measured in a film with $F=0.25$ from temperature $T\sim7$~K to $42$~K. $R_t$ is normalized by the respective zero bias resistance, $R_{t,0}$.}
    \label{Fig_S15}
\end{figure}
\clearpage
\newpage
\section{Discussion on the crossover of spin-relaxation mechanism}
It will be interesting to see if the crossover between the spin-relaxation mechanisms is reflected in the behaviour of the separation between interfaces, the Coulomb screening length, and the spatial extent of the confining potential at the interfaces. However, to rigorously quantify the latter two parameters, we will likely need ab-initio calculations of the structure, which are beyond the scope of the manuscript, as these crucially depend on the band structure, band hybridisation, electric permittivity, effective mass, and so on.
However, below we discuss some possible estimates of these quantities. 
\begin{itemize}
\item{The inter-particle spacing can be approximated by the expression $F/r_{\mathrm{Ag}}^{1/3}$, assuming a periodic distribution of nanoparticles. This is supported by our earlier paper \cite{doi:10.1021/acsaelm.3c00379}, showing the linearity of the measured resistivity with the approximated area of buried interfaces per unit volume based on this expression.}
\item{The Coulomb screening length ($l_{\mathrm{screening}}$) for bulk gold is 0.05 nm. In the Thomas-Fermi approximation of electrostatic screening, this is determined by the permittivity and the charge carrier density. From our low-field Hall measurements, shown in Supplementary Fig.~\ref{Fig_S4}, we do not yet see any appreciable change in the carrier density. However, the permittivity can widely vary with $F$. Indeed, our magnetotransport and tunnelling measurements indicate that the screening, captured by $F_\sigma$, becomes poorer with increasing $F$, becoming the weakest at $F\sim0.12$. We believe that at this value of $F$, where we see the crossover of spin-relaxation mechanism, and rather in a wide region spanning it, $l_{\mathrm{screening}}$ can far exceed the interparticle separation.  Hence, to make a meaningful interpretation, we probably need to look at the spatial extent of the confining potential at the interface.}
\item{The confining potential length at the Ag/Au interface can be approximated by the polaron delocalization length, which can widely vary depending on the specific material properties. For example, this typically ranges between 0.5-1 nm for small polarons in oxide perovskites \cite{PhysRevMaterials.3.114602}, $1-5$~nm for small/large (mixed) polarons in TMDCs \cite{giannini2019quantum}, $1-10$~nm for intermediate/large polarons in organic semiconductors \cite{giannini2019quantum}, $4-10$~nm for large polarons in halide perovskites \cite{doi:10.1021/acs.accounts.1c00675},$3-15$~nm for large polarons in conjugated polymers/crystals \cite{C3CP51477C,C6CP07485E}, 10-100 nm for large/delocalized polarons in inorganic semiconductors \cite{C6CP07485E} and so on. Our temperature-dependent resistivity measurements \cite{kumbhakar2025engineering} have shown that the electrical transport in the Ag@Au system deviates from a conventional metal and can be phenomenologically modelled by a thermally activated parallel channel. This has led us to believe that the Ag/Au interface likely hosts small polarons, which possess a thermally activated hopping mobility. By comparing the polaron delocalization length for small polarons in diverse systems, we assume the same in our system to be around ~1 nm, which is an estimate of the confining potential or charge accumulation at the interface.
}
\end{itemize}

Below in Fig.~\ref{Fig_S16}, we plot the average separation between the adjacent interfaces as a function of $F$. The black solid line represents twice the confining potential depth. The intersection of the curves indicates a region where the conduction electrons of gold are the most likely to experience the surface potential from surrounding interfaces. Hence, within the statistical uncertainty of the size and spatial distribution of the silver nanoparticle, we observe that the crossover of the spin-relaxation mechanisms may correspond to the scale of the interface separation becoming of the order of the localization length at the interface.
\begin{figure}[h!]
    \centering
    \includegraphics[width=0.5\linewidth]{Fig_S16.PNG}
    \caption{Average separation between adjacent interfaces plotted as a function of $F$, illustrating the variation in interfacial spacing with changing $F$.}
    \label{Fig_S16}
\end{figure}
\clearpage
\newpage
\bibliography{bibliography}